# Towards the Framework of the File Systems Performance Evaluation Techniques and the Taxonomy of Replay Traces

Dr. Brijender Kahanwal*
Department of Computer Sc. & Engg.,
GGGI, DINARPUR, AMBALA,
HARYANA
(INDIA).

Dr. Tejinder Pal Singh
Department of Applied Sciences,
RPIIT, BASTARA, KARNAL,
HARYANA
(INDIA)

*Abstract:* This is the era of High Performance Computing (HPC). There is a great demand of the best performance evaluation techniques for the file and storage systems. The task of evaluation is both necessary and hard. It gives in depth analysis of the target system and that becomes the decision points for the users. That is also helpful for the inventors or developers to find out the bottleneck in their systems.

In this paper many performance evaluation techniques are described for file and storage system evaluation and the main stress is given on the important one that is replay traces. A survey has been done for the performance evaluation techniques used by the researchers and on the replay traces. And the taxonomy of the replay traces is described. The some of the popular replay traces are just like, Tracefs [1], //Trace [2], Replayfs [3] and VFS Interceptor [12]. At last we have concluded all the features that must be considered when we are going to develop the new tool for the replay traces. The complete work of this paper shows that the storage system developers must care about all the techniques which can be used for the performance evaluation of the file systems. So they can develop highly efficient future file and storage systems.

*Keyboards:* Performance Evaluation Framework; File Systems; Replay Traces Taxonomy; Evaluation Techniques.

## I. INTRODUCTION

File and storage system designs are being proposed in a little span of time because there is no robust file system is available which can perform all the functionalities according to the always changing user needs. Every user has their specific needs or demands which are not common at all. One user may ask for the secure file system because he/she has important information that must be protected from the others which are not authorized. Some are demanding for highly portable file systems. Considering all these we have also developed Java File Security System (JFSS) [6]. One user demands for the energy efficient file systems because he/she is using portable devices. Because of such diverse requirements by the users it is very typical to develop a robust file system. There are so many different types of file systems available. The user has to choose one of them which are suitable for them. Here the question is which one is better for the selection? To make this judgment we require the evaluation tools. These tools are to be applied by the researchers on the file systems under study for the performance evaluation.

There are so many performance evaluation techniques have been used by the researchers. These are benchmarking, tracing, profile, indirect and ad hoc and energy efficiency measuring techniques. Every technique has an extra overhead on the system performance but with the help of these we can find out the bottlenecks in our file systems and make the regarding performance improvements. The user can also easily find the good product which is a suitable match for his needs. In this paper we have also shown the taxonomy of replay traces according to their features.

The rest of the paper is organized as follows. Section 2 considers the related work. Section 3 provides the hierarchical diagram of the performance evaluation techniques. Section 4 is about the benchmarking, its types, and taxonomy of replay traces are described in more detail. The comparison of the tracing tools is done on the basis of their features. Section 5 contains the descriptions about the profile evaluation technique. Section 6 is about the energy efficient technique for the file systems. Section 7 describes the indirect and ad hoc techniques for the performance evaluation. At last but not least the section 8 about the conclusion and future scope.

## II. RELATED WORK

In this section we give related work in the file system performance evaluation.

Avishay Traeger et al. [7] has done the study near about nine years on the file system benchmarking and they have told so many pros and corns about the benchmarking techniques. According to them it is critical to benchmark when we are evaluating performance and a typical task especially for file and storage systems. In this paper many benchmark tools are discussed. Kester Li [9] has given ideas about the low power file systems. S. Jhou et al. [5] has developed the tracing package for the Berkeley UNIX operating system. D. J. Lilja [10] has discussed so many computer performance evaluation techniques in his book.

The more work is done on especially the tracing tools for the better understanding of the application behavior. So many tracing tools have been studied like Replayfs, Tracefs, //Trace, VFS Interceptor etc. for the taxonomy of replay traces. In all these there is no analysis tool for the processing of the traced records.

### III. PERFORMANCE EVALUATION TECHNIQUES

File System evaluation is most significant part of the research study in file and storage system. It is heavily dependent on the nature of the application generating the load. To achieve optimal performance, the underlying file system configuration must be balanced to match the application characteristics. There are four techniques that are used to evaluate the file systems. These are benchmarking, profiles, energy efficiency techniques and Indirect & ad hoc techniques. The benchmarking is further classified as microbenchmarking, macrobenchmarking and replay traces. All the classification is represented with the help of the figure 3.1. In the next sections they are explained one by one.

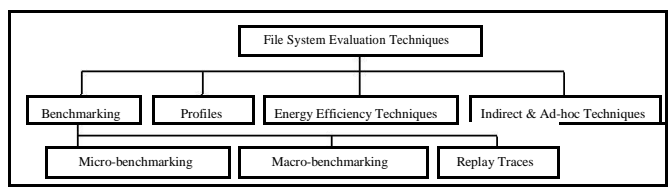

Figure. 3.1 A hierarchical relationship of performance evaluation techniques for file and storage systems.

### IV. BENCHMARKING

Benchmarking is simply the process of measuring the performance. It is critical when we evaluate the performance. Because every system has different features and optimizations, so a single benchmark is not always suitable. There are so many complexities which make benchmarking the systems a challenging task. So many factors which are contributing to the complexity are as follows:

i) *Storage variety*: It is not a single local hard drive on your desktop machine. It is much more like Network-Attached Storage (NAS), Storage Area Network (SAN), flash, RAID etc.
ii) *Various types of File Systems:* There are many types of file systems available. Some are operating on the local machine and using different data structures, logging infrastructures, cryptographic, compression etc. Distributed file system and Network File System behaves differently than the local ones.
iii) *Operating System Variety*: There are so many operating systems with different behaviors. So many operating systems are running on a single machine through virtual machines. Even a single operating system behaves differently on the different configurations.
iv) *The workload*: The user activities and access patterns are difficult to correctly characterize and recreate.
v) *Asynchronous activities*: Other user processes and kernel processes may interact with the storage stack and change the behavior of the system.
vi) *Caches:* Operating system caches are spread at various levels like hard disk cache, buffer cache in
between RAM and processor. The disk caches contain the currently accessed data and metadata which can change the system behavior.

Benchmarks are mostly used to provide an idea of how fast some piece of software or hardware is. This is used by the consumers to take purchasing decisions and is used by the researchers to determine their system. Ideally, the users can test the performance in their own settings or according to their needs using real workloads. All this is impractical and time consuming task to test many systems. Benchmarking the file and storage systems requires complete care which exacerbates the situation.

*A.    The Benchmarking Environment:*

The system state can have a significant effect on results at the time of benchmark execution. We should determine the state. It should be reported with the results correctly. There are few major factors as follows:

*a.    The cache state:*

The cache state of the system may affect the code path which is tested. That will affect the results. The cache has two states either warm or cold. In the real environment the cache is warm and the benchmark that accesses the cached data may be unrealistic. The requests are serviced from the memory and the file system is not exercised properly. The caches should be cleared before benchmarking. It is possible by freeing large amounts of memory, remounting the file system, reloads the storage driver or with the help of rebooting which is effective one.

*b.* **The zoned constant angular velocity of disks:**

The disks are using zoned constant angular velocity (ZCAV) for storing the data. The cylinders are divided into zones where the numbers of sectors in a cylinder are increasing with the distance from the center of the disk. So the transfer rate always varies from zone to zone. This ZCAV effect must be minimized by creating smaller size partitions on the outer part of the disks.

*c.* **The file system aging:**

The benchmarks run on the empty system may produce results differently than the real environment. So it should be aged with the synthetic workloads. We may run long term workloads by copying an existing raw image or to replay a trace before benchmarking.

*d.* **The unnecessary processes during benchmarking:**

All unnecessary processes or services should be stopped before benchmarking. These can affect results of the benchmarking. We should use multithreaded workloads because they correctly show a real system which has many active processes normally.

### B. Types of Benchmarks

There are three types of benchmarks. These are as follows:

*a.* **Macro benchmarks:**

It measures the performance of the file system under some pre-determined workload. It may use either a real application to generate a workload (e.g., compiling and linking a large piece of software), or it may itself be a custom application that drives the file system with a synthetic workload. The goal of macrobenchmark programs to understand how real workloads perform on a file system.

Thus, many macrobenchmarks consist of executing some application with carefully specified parameters. While any program that exercises the file system may be suitable for use as a macrobenchmark, the most common such program is the compiler. Not only does the compiler read and write many files, it is also a tool that researchers frequently use. A typical compilation-based macrobenchmark consists of building and linking the operating system kernel for the system being benchmarked.

*b.* **Micro benchmarks:**

It measure one specific characteristic of file system behavior, such as the time to create or delete a file. With the help of these few operations are exercised. These are used for the better understanding of the macrobenchmark results. These are helpful in isolating the effects of specific parts of the systems and to show the worst case behavior.

*c.* **Replaying Traces:**

These are alternative to benchmarks. They are representative of real applications and are easy to use. File system traces are used for years to analyze user behavior, system software behavior that leads to advances in file system and storage technologies. These are used for user behavior analysis; file system debugging, security & auditing, stress testing, intrusion detection and more recently forensic analysis [7, 14] – to rollback and replay the traced operations or to revert a file system to a state prior to an attack. These have long been used for file system evaluation and optimization. The researchers can replay a file system trace to evaluate a newly designed file system or to find out the bottleneck or special access patterns.

Traces may be collected from a real environment or a synthetic environment. Traces from real environments are most representative and helpful. Trace records are gathered for the file system operations like open, create, close, reads and writes, renames, deletes, executes, forks and exit system calls. We know that the I/O activities are very bursty. The lot of data gathered using the packages and that is important for the studies of disk caches, file migration algorithms, file system performance and load balancing strategies. In the complex system, it is difficult to understand the complete behavior and identify the performance shortcomings and debugging it. A trace driven study or analysis of I/O activities can not reveal the complete behavior of the file system.

**i.** *Design Issues for Tracing File Systems (Tracing Tools):* It is the first step of designing. File systems tracing may be done at different logical levels: block level, the driver level, the Virtual File System (VFS), the network level for Network File System (NFS) or system call level. A tracer can collect records at various degree of granularity like users, groups, processes, files and file names, file operations (open, create, close, reads and writes, renames, deletes, executes, forks, etc.) and more.

We trace the file operations which are performed by the user processes and system processes. It can be done on the top of the Operating System kernel. The user can perform the file operation through system call directly or through the system utility routines. So there are so many places where file operations can be captured. There is always a risk of missing some of them.

Disadvantages of tracing at the Kernel level:

a) The package can't tell from where the file operation comes. That may be through the system call or through the system utility routine. For this we have to trace the command that caused the file operation in parallel and then matching these two records later.
b) The kernel has been changed and debugging it is much tough task as compared to the user program. Any fault in the tracing package may crash the system. For this make minimum changes to the kernel and do the carful implementations.

ii. *Tracing Process:*

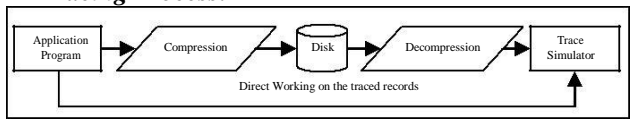

Figure. 4.1 The complete tracing process

The complete tracing process is shown in the above figure. The tracing system has two fundamental components: the application program on which the tracing is done and the trace simulator it processes the trace record data directly or indirectly. Indirectly in the sense the trace records are stored on the large disks and further processed later. The trace records are huge data and they acquire large space and because of that they are compressed to acquire less storage space. The record file of traced data may transformed into aggregate counters, compressed, check summed, encrypted streams. The tracer can buffer and direct the resulting data to various destinations (sockets, disks, etc.). With the help of tracers the research community can devise better software and hardware to accommodate ever-changing computing needs.

iii. **Taxonomy of Replay Traces:** We have constructed a simple taxonomy that captures main features of I/O tracing frameworks which can be utilized by the developers and potential users to formalize their tracing requirements.

Basic requirements for the trace package:

*a)* **Comprehensiveness:** (able to collect all important information): It should provide the clear and detailed snap of the file system activities. We must trace all the relevant file operations, and collect the important information. It must generate complete and accurate information in order to avoid any guessing work. The data should be of wide range to analyze or simulate the studies.

*b)* **Flexibility:** It should be flexible according to the changing trace needs. Tracing may be done for the single user, single process, a group of users or a group of processes. We may be able to activate or deactivate the trace dynamically according to the needs without any disruption in the services of the system. The behavior may be study for the long time or for the short time of a file system.

*c)* **Minimum Performance Consequences:** The tracing overhead penalty must be kept low. When the trace is not activated, there should not be any extra burden on the system which is caused by the tracing package. And when it is activated, it should not degrade the overall system performance.
  i) No extra amount of computing for extracting data by a trace package.
  ii) The amount of data generated should not be large.
  iii) Both aspects are important if the trace is run for several days.

*d)* **Minimum Changes to the System:** It is important to keep minimum tracer code required and that will minimize the degree of maintenance. It will tend to reduce the performance penalty on the system.

*e)* **Convenience for Analysis:** The final format of the trace data should be easily and simply analyzable with only simple computations for generating simple outputs.

*f)* **High availability:** For example, in supercomputing environments, tasks may last for months and if stopped, they must restart from the beginning and this in not affordable.

*g)* **Security:** Traces are used to monitor malicious activities on the system. So, it is important that the generated traces should be protected from attacks or subversion. The encryption and keyed checksums can provide strong security.

*h)* **Portability:** The tracing tool can be stacked above the underlying file system and then that can easily trace. They must be highly portable.

*i)* **Privacy:** Traces can raise concern about user's privacy because they may contain personal information when they are publically distributed. Such type of information can't simply be removed from traces since it required for correlation.

*j)* **Anonymization:** Traces are always collected for the distribution. It is desirable to anonymize personal or sensitive data. It must specify which parts of the trace must be anonymized.

*k)* **Parallel file system compatibility:** The tool must work with other applications in parallel and easily augmented to add parallel functionalities.

*l)* **Control of trace granularity:** The person who is using the trace data must collect the data as much is needed because there is a performance overhead.

*m)* ***Trace data format:*** It may be binary format that is usable or analyzable by the computer only that saves memory space and facilitate automated parsing. Sometimes it is convenient to view by the humans that format is the visually inspected trace formats. They may be compressed, encrypted, or checksumming.

*n)* ***Elapsed time overhead:*** There should have accountability of the time taken by the trace tool that is completely performance overhead.

*o)* ***Analysis tools:*** It is most important part of any tracing package that the analysis tool should manipulate the traced records and does analysis on that to produce the important results on which decisions can be made.

*p)* ***Ease of installation &use:*** The installation of the tracer package should be painless. The traced records should be simple to use for further interpretations.

*q)* ***Traced event types:*** Various types of events like I/O function calls, system calls and file system operations might be traced.

*r)* ***Compression of traces:*** The traced records are recorded in the compressed form on the disks.

The above mentioned requirements often conflicts with each other, but these can be reduced by careful design and implementations.

Table 4.1 Comparison of various traces according to some specified features

| Sr. No. | Features | *Tracefs* (A Stackable File System) | *//TRACE* | *Replayfs* | *VFS Interceptor* |
|---|---|---|---|---|---|
| 1 | Traced event types | File System Operations | I/O system calls | File System Operations | File System Operations |
| 2 | Ease of installation & use | Typical | Easy | Typical | Easy |
| 3 | Trace data format | Binary | Human-readable | Binary | Binary |
| 4 | Analysis Tool | No | No | No | No |
| 5 | Anonymization | No | No | No | No |
| 6 | Compression of Traces | Yes | No | No | No |
| 7 | Parallel file system compatibility | No | Yes | No | No |
| 8 | Control of trace granularity | Yes | No | No | No |

In conclusion, a file system benchmark should highlight the high-level as well as low-level performance. For the high-level view of performance measure, we may use at least one macrobenchmark and for the low-level view of performance measure, we may use many microbenchmarks. We should consider the following benchmark properties:

i) The benchmarks may be CPU bound or I/O bound. For the file systems they should be I/O bound.
ii) It should record accurate measurements for timings.
iii) It should be scalable means exercise every machine.
iv) It should be independent of hardware or software speeds.
v) It should record multithreaded workloads which provide more realistic views.
vi) The outputs generated by them should be well understood.
vii) They should be portable.

## V. PROFILING

It gives the overall view of the execution behavior of the software under study. It measures about how much time or the fraction of total time, the software is spending on assured states. It also shows the subroutine-oriented execution time for identifying the portions of the software which are consuming largest amount of time fraction from the total time. The system-level performance bottlenecks are identified. So the developers can enhance the performance of the system. It is a periodic sampling of a program's execution. It obtains average program performance for cache misses, clocks per instruction etc. We can create profiles with the help of two different techniques – program counter (PC) sampling and block-based counting.

A very few or less number of developed tools are available to profile file system performance. They are highly dependent on the workloads. Disk operations include mechanical latencies to position the head
. The longest

operation is seeking, or moving the head from one track to another. Therefore, file systems are designed to avoid seeks. Unfortunately, modern hard drives expose little information about the drive's internal data placement. The OS generally assumes that blocks with close logical block numbers are also physically close to each other on the disk. Only the disk drive itself can schedule the requests in an optimal way and only the disk drive has statistical information about its internal operations.

### A. *Program Counter (PC) based Profiles:*

It relies on periodic interrupts during a program's execution to sample the PC. It is a statistical measurement technique in which a subset of the members of a population is being examined randomly. The profiled program does not know when to be interrupted in random sampling. The sampling is done for the long period, because it can accurately reflect the program activity. This type of profiling systems takes a snapshot of kernel structures to provide an average profile.

### B. *Block-based Counting Profiles:*

It is an alternative approach to produce an exact execution profile by counting the number of times each basic block is executed. The structure of the basic block can be changed to generate a profile by inserting additional instructions. These are used for counting the time for the executed block. After the termination of the program execution, a histogram is formed with the help of these values for each block's frequency of execution. This histogram shows which portion of the program is executed most.

The main difference between these two is that the block-based profiles show the exact picture of the execution frequencies. But the PC counter based profile is only a statistical technique which estimates the frequencies of the software. The block-based profiles provide a much more run time overhead for the evaluation.

## VI. ENERGY EFFICIENCY TECHNIQUES

Power efficiency has become a major concern for computing systems. The performance of the file systems may be measured in the terms of power consumption. The fundamental goal of the energy efficient file system design is saving energy without sacrificing performance. The energy consumed by the hard disk drive is shown with the help of the following equation:

$$E_{HDD} = E_S + E_H \quad (1)$$

Where $E_{HDD}$ refers to the total energy consumed by the hard disk drive, $E_S$ refers to the energy consumed by the spinning of the platter, and $E_H$ refers to the energy consumed by the head movement.

These are the necessary parts for a low power consumption file systems [10] for the portable devices:
i) Fine-grained disk spindown
ii) Whole file pre-fetching cache
iii) 8-16 MB of low power, low read latency memory

If the system will miss any one of these three requirements then there will be a tradeoff between power and performance of it. Without fine-grained disk spindown, file system power consumption will be high with good performance. Without a whole file pre-fetching cache, the choice is between coarse-grained disk spindown with high power consumption and good performance or fine-grained disk spindown with low power consumption and poor performance. Without a low power memory, the cache may consume same power as much by disk spinning. So they collectively make low power file system.

The energy efficient file systems must follow these basic guidelines: delayed updates, aggressive prefetching, and urgency based scheduling, compression and device awareness.

## VII. INDIRECT & AD HOC BENCHMARKS

At many times, we have no direct performance metric for the evaluation. So we develop or go for the indirect ways to measure just like the Ad Hoc technique. We drive the results indirectly. We have no direct measurement tool for the energy consumption of file systems, and then we drive the results taken by other methods. For example, suppose we are not able to measure the desired quantity directly, but we may be able to measure another related value directly. Then we deduce the desired values from the other measured values. These are the benchmarks which are written by the authors for in house use. These are not widely used. These are not tested as much as the widely used benchmarks, so these are very prone to errors. One good thing about these is that they are well understood by the researchers who have developed them. The early use of these is helpful for the optimization.

## VIII. CONCLUSIONS AND FUTURE SCOPE

The goal of this work is to provide an insight into the performance evaluation techniques which may be helpful for the researchers and the users of file and storage systems. The more stress is given on basically on the benchmarking techniques especially replay traces. We have also provided the guidelines for the researchers to develop a good tracing tool which has the good features. After that we have also compared three traces on the basis of features. Any one of them has not provided the analysis tools for the gathered information. The replay traces explains the complete behavior of the application program or system software. But that is a time consuming process to reach to any conclusions.

We intend to develop a tracing file system which will include all the basic functionalities and analysis tool with it.